# Experimenting with Generative AI:
# Does ChatGPT Really Increase Everyone's Productivity?

Voraprapa Nakavachara, Tanapong Potipiti[1], Thanee Chaiwat
Faculty of Economics, Chulalongkorn University

29 February 2024

Generative AI technologies such as ChatGPT, Gemini, and MidJourney have made remarkable progress in recent years. Recent literature has documented ChatGPT's positive impact on productivity in areas where it has strong expertise—attributable to extensive training datasets—such as the English language and Python/SQL programming. However, there is still limited literature regarding ChatGPT's performance in areas where its capabilities could still be further enhanced. This paper aims to fill this gap. We conducted an experiment in which economics students were asked to perform writing analysis tasks in a non-English language (specifically, Thai) and math & data analysis tasks using a less frequently used programming package (specifically, Stata). The findings suggest that, on average, participants performed better using ChatGPT in terms of scores and time taken to complete the tasks. However, a detailed examination reveals that 34% of participants saw no improvement in writing analysis tasks, and 42% did not improve in math & data analysis tasks when employing ChatGPT. Further investigation indicated that higher-ability students, as proxied by their econometrics grades, were the ones who performed worse in writing analysis tasks when using ChatGPT. We also found evidence that students with better digital skills performed better with ChatGPT. This research provides insights on the impact of generative AI. Thus, stakeholders can make informed decisions to implement appropriate policy frameworks or redesign educational systems. It also highlights the critical role of human skills in addressing and complementing the limitations of technology.

---

[1] Corresponding Author: tanapong.p@chula.ac.th

We are grateful for the funding provided by the Puey Ungphakorn Institute for Economic Research at the Bank of Thailand. We also extend our sincere appreciation for the research assistance provided by Chanalak Chaisrilak and Nathatai Kullapa-anunchan. The research conducted in this paper was approved by Chulalongkorn University's Research Ethics Review Committee for Research Involving Human Subjects.

LLM Assistant Statement: This article was copy-edited with the assistance of GPT-4.

1. Introduction

Generative AI technologies such as ChatGPT, Gemini, and MidJourney are experiencing rapid advancements and are anticipated to improve significantly in the coming years. Currently, they have the capability to execute a diverse array of tasks, including but not limited to composing text, generating ideas, writing codes, and creating artwork. Major industry players, such as Microsoft/OpenAI and Google, are channeling substantial investments into these technologies, ensuring their evolution.

Recent literature has begun to explore how these Generative AI technologies can impact labor productivity and the economy. Brynjolfsson et al. (2023) asked customer support agents to use AI-enabled tools and found a 14 percent improvement in their productivity. Noy and Zhang (2023) conducted an experiment with mid-level professional writing tasks and observed that participants using ChatGPT completed the tasks faster and with higher quality. For tasks related to economic research tasks, Korinek (2023) noted that ChatGPT can be useful for a variety of tasks, including ideation and feedback, writing, background research, coding, data analysis, and mathematics. Cheng et al. (2023) explored whether GPT-4 could serve as an effective data analyst. The authors found that GPT-4 can outperform an entry-level data analyst in terms of efficiency and cost, while also delivering results more quickly.

Most of the literature has documented ChatGPT's positive impact on productivity, particularly in tasks that fall within its areas of strong expertise—attributable to extensive training datasets—such as the English language and Python/SQL programming. However, there appears to be a gap in the literature concerning ChatGPT's performance in areas where its capabilities could be further enhanced. This paper aims to address this gap.

In this paper, we conducted an experiment in which economics students were asked to perform writing analysis tasks in a non-English language (specifically, Thai) and math & data analysis tasks using a less frequently used programming package (specifically, Stata). To the best of our knowledge, this paper is among the first to conduct a ChatGPT experiment in Non-English and Non-Python/SQL environments, areas in which ChatGPT still has room for improvement.

The results indicate that, on average, students demonstrated improved performance in both scores and time taken to complete the tasks when utilizing ChatGPT. Yet, a closer analysis unveils that 34% of participants saw no improvement in writing analysis tasks, and 42% did not improve in math & data analysis tasks when employing ChatGPT. Further scrutiny revealed that higher-ability students, as proxied by their econometrics grades, were the ones who performed worse in writing analysis when using ChatGPT. Additionally, our findings suggest that students possessing advanced digital skills experienced improved performance when using ChatGPT.

This research offers valuable insights for educators, policymakers, businesses, and workers, enabling them to comprehend and predict the ways in which generative AI technologies may enhance or diminish the performance of students and employees. Consequently, stakeholders can make informed decisions to adapt strategies or implement appropriate policy frameworks. Moreover, the study contributes to the education sector by offering guidance on how to redesign educational systems in response to the advent of generative AI. A critical takeaway is the enduring importance of human skills in recognizing the limitations of these technologies and compensating for them, highlighting the synergy between human expertise and artificial intelligence.

## 2. Literature Review

The impact of technological advancements on the economy is a well-explored theme within economic literature. While some economists express concerns over potential worker displacement (Frey & Osborne, 2017; Lekfuangfu & Nakavachara, 2021), others, in line with the goals of this research, focus on how these advancements can have an impact on productivity.

Generative AI, particularly ChatGPT, has gained popularity among the public and also in economics literature. In the study conducted by Brynjolfsson et al. (2023), about 5,179 customer support agents were involved. Approximately half of them were allowed to use tools linked to ChatGPT, while the rest were not allowed. The researchers found that the group using ChatGPT experienced a 14 percent improvement in productivity, as measured by the number of cases solved per hour. In another paper by Noy and Zhang (2023), an experiment was conducted involving 444 workers engaged in mid-level professional writing tasks. The researchers discovered that the group using ChatGPT experienced a decrease in the time taken by 0.8 standard deviations, while the output quality increased by 0.4 standard deviations. Both studies pointed out that ChatGPT helped improve the productivity of the inexperienced more than that of the experienced individuals.

In the realm of risk analysis, Kim et al. (2023) applied ChatGPT to evaluate firm risk from quarterly earnings call transcripts. The authors revealed that ChatGPT's risk assessment measure has a positive correlation with stock price volatility. Korinek (2023) highlighted that ChatGPT could be beneficial for a variety of economic research tasks, including ideation and feedback, writing, background research, coding, data analysis, and mathematics, although its effectiveness may vary depending on the specific task.

Dell'Acqua et al. (2023) carried out an experiment involving consultants from Boston Consulting Group, categorizing them into three groups: those not permitted to use AI, those using GPT-4, and those using GPT-4 AI with an overview. For tasks within AI's strengths, such as generating and developing new product ideas, consultants utilizing AI demonstrated notable increases in productivity. Conversely, for tasks considered outside AI's capabilities, including solving business problems with quantitative data and performing customer and company interviews, consultants employing AI performed slightly worse than their counterparts without AI.

Cheng et al. (2023) investigated the potential of GPT-4 as a competent data analyst. They discovered that GPT-4 surpasses the performance of an entry-level data analyst in efficiency and cost-effectiveness, also providing faster outcomes.

Choi and Schwarcz (2024) conducted a study where legal students used ChatGPT during their exams. They observed that GPT-4 significantly improved performance in straightforward multiple-choice questions but had little effect on complex essay questions. Furthermore, the impact of GPT-4 varied greatly based on the students' initial skill levels; those with lower starting points experienced substantial improvements with AI help, whereas top-performing students encountered declines in performance.

## 3. Experimental Design

We conduct an experiment to examine how ChatGPT affects student performance in writing analysis tasks in Thai and math & data analysis tasks in Stata. This study involved 121 college economics students from Chulalongkorn University and Thammasat University in Thailand. The writing analysis tasks encompassed brainstorming, reading and providing feedback on texts, composing tweets, and summarizing documents. The math & data analysis tasks included coding for data visualization, variable generation, conducting regression analyses, hypothesis testing, and equation derivation.

Participants began with two sessions of writing analysis tasks, comparing performance with and without ChatGPT. They then proceeded to two sessions of math & data analysis tasks, again with and without ChatGPT. In all sessions, they had access to a browser/internet but were not allowed to use other Large Language Model (LLM) platforms. To accurately assess the impact of ChatGPT, we prepared two sets of tasks (Set A and Set B) of equal difficulty for both writing analysis and math & data analysis. This setup allows for a direct comparison of individual performance with versus without ChatGPT. To eliminate any bias from the difference in problem sets or the order of using ChatGPT, participants were randomly divided into four groups as detailed in Table 1. For instance, Group 1 started with Set A of writing analysis without ChatGPT, followed by Set B with ChatGPT, then proceeded to Set A of math & data analysis without ChatGPT, and finished with Set B with ChatGPT. The sequence for the other groups is outlined in Table 1.

We assessed performance using two metrics: a score out of 10 (averaged from the assessments of three experts) and the time required to complete tasks. Participants were awarded monetary rewards based on their score (with higher scores preferred) and completion time (with quicker times preferred). Each student received a base show-up fee plus a performance fee determined by both the time spent and the quality of their work. Each task session had a 20-minute limit. Participants could submit their work before this time, but were required to submit whatever they had at the 20-minute mark. Following all sessions, participants completed a questionnaire detailing their academic background (including GPAX and econometrics grade) and self-assessed their reading, writing, math, and digital skills on a scale from 1 to 5.

## 4. Data Overview and Preliminary Results

Table 2 provides a summary statistics for the 121 students participating in the study. Most them are in their 3rd and 4th years. Their average Grade Point Average (GPAX) is 3.32, with an average econometrics grade of 3.00. About 36% of these students are male. They rated their abilities in reading, writing, math, and digital skills on a scale from 1 to 5 (where 5 is the highest), with average scores of 3.46, 3.07, 3.26, and 3.20, respectively. Notably, approximately 6.6% of the participants reported never having used ChatGPT. In preparation for our main regression analysis, detailed in Section 5, we defined a 'ChatGPT proficiency' variable based on 'ChatGPT usage per week,' with thresholds at 30 minutes, 1 hour, and 2 hours to categorize three levels of expertise with ChatGPT.

Table 3 revealed our initial findings. In Panel A, it can be observed that, on average, participants achieved higher scores when permitted to use ChatGPT for both writing analysis and math & data analysis tasks. Panel B illustrated that, on average, participants completed tasks, both writing analysis and math & data analysis, more quickly when using ChatGPT.

Figure 1 displayed the score and time distribution for both sets of tasks. For scores, the distribution for participants using ChatGPT lies on the right, indicating higher scores compared to those not using ChatGPT. Conversely, for time, the distribution for participants using ChatGPT lies towards the left, suggesting they completed tasks more swiftly than those not using ChatGPT.

Table 4 delves deeper into the analysis of scores for writing tasks. 'Score Diff' is defined as the score a student achieved when using ChatGPT minus the score achieved when not using ChatGPT. On the other hand, 'Time Diff' is defined as the time taken when not using ChatGPT minus the time taken when using ChatGPT. On average, the score with ChatGPT is higher than the score without ChatGPT by 0.43 (out of a total possible score of 10), while the time to complete tasks with ChatGPT is shorter than without by 1.73 minutes. Despite the overall improvement in scores with ChatGPT, 41 out of 121 students (approximately 34%) did not see improved outcomes. This discrepancy underscores the importance of investigating the characteristics that differentiate students who benefited from ChatGPT from those who did not. Interestingly, students who did not benefit from ChatGPT generally have higher GPAX and econometrics grades. They also self-report higher reading and writing skills. In contrast, students who benefited from ChatGPT exhibit stronger math and digital skills. In testing for differences in means, only the Econometrics grade showed statistical significance at the 10% level. Figure 2 explores this issue from a distributional angle, contrasting students who improved with ChatGPT against those who did not. Panel B shows that the distribution of econometrics grades among students who did not benefit from ChatGPT tends to lie to the right when compared to their counterparts. Patterns GPAX and for other skills, however, remain inconclusive.

Table 5 looks into the details of math & data analysis scores. On average, scores with ChatGPT are higher than those without ChatGPT by 1.65 (out of a total score of 10), while the time taken with ChatGPT is shorter than without by 0.99 minutes. Despite the general improvement in scores with ChatGPT, 51 out of 121 students (approximately 42%) did not perform better. Further observation reveals that, on average, students who did not improve with ChatGPT tend to have higher GPAX, econometrics grades, and self-assessed reading and writing skills. Conversely, students who improved with ChatGPT demonstrate better math and digital skills. However, none of the t-tests reached statistical significance. Figure 3 explores this issue through the distribution of scores, but the patterns remain inconclusive.

## 5. Empirical Models and Results

To investigate the factors contributing to individuals' performance improvement with ChatGPT, we further employ Ordinary Least Squares (OLS) regression analysis. The main analysis is conducted using the following OLS model.

$$Score\_Diff_i = \alpha + \beta_1 ChatExp_i + \beta_2 Econo_i + \beta_3 GPAX_i + \gamma \cdot Skills_i + \delta \cdot x_i + \varepsilon_i \quad (1)$$

As previously defined in the preceding section, *Score_Diff* refers to the difference in scores a student achieved when using ChatGPT versus not using ChatGPT. *ChatExp* is a dummy variable indicating 'ChatGPT proficiency,' determined by whether participants possess expertise in using ChatGPT. This expertise is assessed based on 'ChatGPT usage per week,' with thresholds at 30 minutes, 1 hour, and 2 hours to categorize three levels of proficiency with ChatGPT. *Econo* represents the student's grade in econometrics. *GPAX* denotes the cumulative grade point average. *Skills* is a vector comprising self-evaluated skills in four domains: reading, writing, math, and digital. *x* represents a vector of other control variables, including a male dummy, group dummies, and college year dummies. This analysis is conducted separately for writing analysis tasks and math & data analysis tasks. We employ robust standard errors in all our regressions to ensure the reliability of our findings.

In the supplemental analyses, we modified the dependent variable, *Score_Diff*, by substituting it with a dummy variable set to 1 for students who performed better with ChatGPT and 0 for those who did not. We applied Equation (1) using the Linear Probability Model (LPM) for both writing analysis tasks and math & data analysis tasks. Additionally, we conducted the analysis using a logistic regression model for both sets of tasks to compare outcomes. Robust standard errors were utilized in all regressions to ensure the accuracy and reliability of the results.

Table 6 presents our main regression findings. Columns 1-4 display results for the writing analysis tasks. Column 1 excludes the *ChatExp* variable, while Columns 2-4 incorporate *ChatExp* with thresholds of 30 minutes or more, 1 hour or more, and 2 hours or more per week, respectively. The *Econo* variable is consistently negative and significant across all specifications at the 10% level, suggesting that students with higher econometrics grades performed worse when using ChatGPT, with scores approximately 0.54 to 0.57 points lower. Similarly, Columns 5-8 detail results for the math & data analysis tasks. Column 5 omits the *ChatExp* variable, whereas Columns 6-8 include *ChatExp* at thresholds of 30 minutes or more, 1 hour or more, and 2 hours or more per week, respectively. The *Digital skills* variable is positive and significant at the 10% level in all models, except for Column 6, implying that students with higher digital skills scores performed better when using ChatGPT, with an approximate increase of 0.70 to 0.73 points in their scores.

Table 7 provides the supplemental regression results utilizing the Linear Probability Model (LPM). Columns 1-4 display the outcomes for writing analysis tasks. Consistent with previous findings, the *Econo* variable is negative and significant at the 5% level across all specifications, indicating that students with higher econometrics grades performed worse when using ChatGPT. Similarly, Columns 5-8 present results for the math & data analysis tasks. Once again, the *Digital Skills* variable is positive and significant at the 5% level in all specifications, demonstrating that students with superior digital skills achieved better results with the use of ChatGPT.

Table 8 showcases the supplemental regression outcomes employing the logistic model. In the analysis of writing tasks, presented in Columns 1-4, the *Econo* variable consistently appears negative and significant at the 5% level across all specifications. This pattern suggests that students with higher econometrics grades tend to perform worse when utilizing ChatGPT. Additionally, the *Digital Skills* variable is positive and significant at the 10% level in all specifications, indicating that students with superior digital skills achieve better results with ChatGPT. For the math & data analysis tasks, detailed in Columns 5-8, the *Digital Skills* variable again shows a positive and significant relationship, at levels ranging from 5% to 10%, across all specifications. This reaffirms that students with enhanced digital skills tend to excel when using ChatGPT.

## 6. Conclusion

In this paper, we explore domains where ChatGPT's expertise may not yet be fully developed, specifically focusing on tasks conducted in a non-English language (Thai) and math & data analysis tasks utilizing a less commonly used programming package (Stata). We conducted an experiment with 121 economics students, asking them to complete writing analysis tasks in Thai and math & data analysis tasks in Stata. Our findings indicate that, on average, participants achieved better scores and completed tasks more quickly using ChatGPT. However, a closer examination reveals that 34% and 42% of participants did not demonstrate improvement with ChatGPT for writing analysis and math & data analysis tasks, respectively. Further analysis showed that students with higher econometrics grades—serving as a proxy for higher ability—tended to perform worse in writing analysis tasks when using ChatGPT. Conversely, students with superior digital skills were found to perform better with ChatGPT across tasks.

Interestingly, our model showed that the *GPAX* variable was not significant in any of our models, whereas econometrics grades did. Our hypothesis is that econometrics grades might serve as a more accurate measure of economics students' abilities compared to GPAX, which encompasses a broader range of subjects, including those not directly related to economics (e.g., General Education). Additionally, the *ChatExp* variable, a proxy of 'ChatGPT proficiency' determined by how oftern each student uses ChatGPT per week, was never significant our models.

On the other hand, the *digital skills* variable was significant. Our hypothesis is that general digital competencies may enable students to more effectively engage with any digital or technical program, including ChatGPT, regardless of their frequency of ChatGPT usage per week.

This research provides critical insights for educators, policymakers, business leaders, and workers, facilitating a deeper understanding of how generative AI technologies can either augment or impair student and employee performance. As a result, stakeholders are better equipped to make well-informed decisions, adapt their strategies, or develop suitable policy frameworks in response. Furthermore, this study makes a significant contribution to the field of education by offering recommendations on how educational systems might be restructured in light of generative AI's emergence. A key takeaway is the pivotal role of human skills in identifying and mitigating the limitations of these technologies, underscoring the complementary relationship between human expertise and artificial intelligence. This highlights the necessity for ongoing education and skill development in maximizing the benefits of AI while addressing its challenges.

Table 1: Experimental Design

|  | Writing Analysis Tasks | | Math & Data Analysis Tasks | |
| --- | --- | --- | --- | --- |
|  | Session I | Session II | Session I | Session II |
| Group 1 (21 Participants) | Set A No ChatGPT | Set B With ChatGPT | Set A No ChatGPT | Set B With ChatGPT |
| Group 2 (34 Participants) | Set A With ChatGPT | Set B No ChatGPT | Set A With ChatGPT | Set B No ChatGPT |
| Group 3 (44 Participants) | Set B No ChatGPT | Set A With ChatGPT | Set B No ChatGPT | Set A With ChatGPT |
| Group 4 (22 Participants) | Set B With ChatGPT | Set A No ChatGPT | Set B With ChatGPT | Set A No ChatGPT |

Table 2: Summary Statistics

Panel A: Overall Statistics

| Variable | Obs | Mean | Std. Dev. | Min | Max |
|---|---|---|---|---|---|
| Male | 121 | 0.36 | 0.48 | 0.00 | 1.00 |
| CollegeYear | 121 | 3.28 | 0.49 | 2.00 | 5.00 |
| GPAX | 121 | 3.32 | 0.33 | 2.50 | 3.99 |
| Econometrics Grade | 121 | 3.00 | 0.78 | 1.00 | 4.00 |
| Self-Evaluated Reading Skills | 121 | 3.46 | 0.82 | 1.00 | 5.00 |
| Self-Evaluated Writing Skills | 121 | 3.07 | 0.85 | 1.00 | 5.00 |
| Self-Evaluated Math Skills | 121 | 3.26 | 1.01 | 1.00 | 5.00 |
| Self-Evaluated Digital Skills | 121 | 3.20 | 0.97 | 1.00 | 5.00 |

Panel B: Breakdown of ChatGPT Usage per Week

| Variable | Obs | Percent |
|---|---|---|
| ChatGPT Usage per Week | 121 | 100.00 |
| Never Used ChatGPT | 8 | 6.61 |
| Up to <15 Minutes | 27 | 22.31 |
| 15 to <30 Minutes | 28 | 23.14 |
| 30 to <60 Minutes | 25 | 20.66 |
| 1 to <2 Hours | 15 | 12.40 |
| 2 to <4 Hours | 8 | 6.61 |
| 4 Hours or More | 10 | 8.26 |

Table 3: Preliminary Results

Panel A: Score (Maximum is 10)

| Variable | Obs | Mean | Std. Dev. | Min | Max |
|---|---|---|---|---|---|
| Writing Analysis - No ChatGPT | 121 | 5.33 | 1.47 | 1.42 | 8.92 |
| Writing Analysis - With ChatGPT | 121 | 5.75 | 1.47 | 1.42 | 8.75 |
| Math & Data Analysis - No ChatGPT | 121 | 3.09 | 2.91 | 0.00 | 10.00 |
| Math & Data Analysis - With ChatGPT | 121 | 4.74 | 2.97 | 0.00 | 10.00 |

Panel B: Time Spent (Maximum is 20 Minutes)

| Variable | Obs | Mean | Std. Dev. | Min | Max |
|---|---|---|---|---|---|
| Writing Analysis - No ChatGPT | 121 | 19.19 | 1.62 | 9.00 | 20.00 |
| Writing Analysis - With ChatGPT | 121 | 17.46 | 3.09 | 7.00 | 20.00 |
| Math & Data Analysis - No ChatGPT | 121 | 18.26 | 2.57 | 5.00 | 20.00 |
| Math & Data Analysis - With ChatGPT | 121 | 17.27 | 3.04 | 6.00 | 20.00 |

Table 4: Detailed Results -- Writing Analysis

Panel A: Overall Difference

| Variable | Obs | Mean | Std. Dev. | Min | Max |
|---|---|---|---|---|---|
| Score Diff = With ChatGPT - No ChatGPT | 121 | 0.43 | 1.98 | -5.00 | 5.17 |
| Time Diff = No ChatGPT - With ChatGPT | 121 | 1.73 | 3.43 | -9.00 | 13.00 |

Panel B: Students with Score Diff > 0

| Variable | Obs | Mean | Std. Dev. | Min | Max |
|---|---|---|---|---|---|
| Male | 80 | 0.40 | 0.49 | 0.00 | 1.00 |
| CollegeYear | 80 | 3.34 | 0.53 | 2.00 | 5.00 |
| GPAX | 80 | 3.31 | 0.34 | 2.53 | 3.99 |
| Econometrics Grade | 80 | 2.91 | 0.78 | 1.00 | 4.00 |
| Self-Evaluated Reading Skills | 80 | 3.41 | 0.82 | 1.00 | 5.00 |
| Self-Evaluated Writing Skills | 80 | 3.05 | 0.91 | 1.00 | 5.00 |
| Self-Evaluated Math Skills | 80 | 3.29 | 1.03 | 1.00 | 5.00 |
| Self-Evaluated Digital Skills | 80 | 3.23 | 0.97 | 1.00 | 5.00 |

Panel C: Students with Score Diff <= 0

| Variable | Obs | Mean | Std. Dev. | Min | Max |
|---|---|---|---|---|---|
| Male | 41 | 0.29 | 0.46 | 0.00 | 1.00 |
| CollegeYear | 41 | 3.17 | 0.38 | 3.00 | 4.00 |
| GPAX | 41 | 3.34 | 0.30 | 2.50 | 3.85 |
| Econometrics Grade | 41 | 3.17 | 0.76 | 1.50 | 4.00 |
| Self-Evaluated Reading Skills | 41 | 3.56 | 0.81 | 2.00 | 5.00 |
| Self-Evaluated Writing Skills | 41 | 3.10 | 0.74 | 2.00 | 5.00 |
| Self-Evaluated Math Skills | 41 | 3.20 | 0.98 | 1.00 | 5.00 |
| Self-Evaluated Digital Skills | 41 | 3.15 | 0.99 | 1.00 | 5.00 |

Table 5: Detailed Results -- Math & Data Analysis

Panel A: Overall Difference

| Variable | Obs | Mean | Std. Dev. | Min | Max |
|---|---|---|---|---|---|
| Score Diff = With ChatGPT - No ChatGPT | 121 | 1.65 | 3.67 | -7.50 | 8.75 |
| Time Diff = No ChatGPT - With ChatGPT | 121 | 0.99 | 2.90 | -7.00 | 10.00 |

Panel B: Students with Score Diff > 0

| Variable | Obs | Mean | Std. Dev. | Min | Max |
|---|---|---|---|---|---|
| Male | 70 | 0.36 | 0.48 | 0.00 | 1.00 |
| CollegeYear | 70 | 3.29 | 0.51 | 2.00 | 5.00 |
| GPAX | 70 | 3.31 | 0.33 | 2.50 | 3.99 |
| Econometrics Grade | 70 | 2.96 | 0.80 | 1.00 | 4.00 |
| Self-Evaluated Reading Skills | 70 | 3.43 | 0.84 | 1.00 | 5.00 |
| Self-Evaluated Writing Skills | 70 | 3.03 | 0.85 | 1.00 | 5.00 |
| Self-Evaluated Math Skills | 70 | 3.26 | 0.93 | 1.00 | 5.00 |
| Self-Evaluated Digital Skills | 70 | 3.30 | 0.86 | 1.00 | 5.00 |

Panel C: Students with Score Diff <= 0

| Variable | Obs | Mean | Std. Dev. | Min | Max |
|---|---|---|---|---|---|
| Male | 51 | 0.37 | 0.49 | 0.00 | 1.00 |
| CollegeYear | 51 | 3.27 | 0.45 | 3.00 | 4.00 |
| GPAX | 51 | 3.33 | 0.32 | 2.73 | 3.93 |
| Econometrics Grade | 51 | 3.04 | 0.76 | 1.50 | 4.00 |
| Self-Evaluated Reading Skills | 51 | 3.51 | 0.78 | 2.00 | 5.00 |
| Self-Evaluated Writing Skills | 51 | 3.12 | 0.86 | 2.00 | 5.00 |
| Self-Evaluated Math Skills | 51 | 3.25 | 1.13 | 1.00 | 5.00 |
| Self-Evaluated Digital Skills | 51 | 3.06 | 1.10 | 1.00 | 5.00 |

Table 6: Main Regression Results

| VARIABLES | (1) Writing Analysis Score Diff | (2) Writing Analysis Score Diff | (3) Writing Analysis Score Diff | (4) Writing Analysis Score Diff | (5) Math & Data Analysis Score Diff | (6) Math & Data Analysis Score Diff | (7) Math & Data Analysis Score Diff | (8) Math & Data Analysis Score Diff |
|---|---|---|---|---|---|---|---|---|
| ChatExpert30m |  | -0.0297 |  |  |  | 0.486 |  |  |
|  |  | (0.349) |  |  |  | (0.628) |  |  |
| ChatExpert1hr |  |  | -0.211 |  |  |  | -0.207 |  |
|  |  |  | (0.375) |  |  |  | (0.747) |  |
| ChatExpert2hr |  |  |  | -0.448 |  |  |  | -1.206 |
|  |  |  |  | (0.437) |  |  |  | (1.025) |
| Male | 0.446 | 0.444 | 0.435 | 0.459 | -0.351 | -0.319 | -0.362 | -0.317 |
|  | (0.410) | (0.414) | (0.416) | (0.417) | (0.793) | (0.796) | (0.795) | (0.786) |
| Econometrics Grade | -0.538* | -0.536* | -0.543* | -0.573* | -0.645 | -0.674 | -0.650 | -0.738 |
|  | (0.296) | (0.297) | (0.300) | (0.304) | (0.575) | (0.578) | (0.578) | (0.598) |
| GPAX | 0.849 | 0.842 | 0.869 | 0.882 | -0.160 | -0.0451 | -0.140 | -0.0691 |
|  | (0.735) | (0.734) | (0.739) | (0.740) | (1.146) | (1.155) | (1.162) | (1.128) |
| Self-Evaluated Reading Skills | -0.201 | -0.203 | -0.213 | -0.210 | 0.0992 | 0.131 | 0.0870 | 0.0726 |
|  | (0.255) | (0.252) | (0.257) | (0.254) | (0.451) | (0.450) | (0.460) | (0.457) |
| Self-Evaluated Writing Skills | 0.0341 | 0.0330 | 0.0364 | 0.0228 | 0.187 | 0.204 | 0.189 | 0.156 |
|  | (0.248) | (0.252) | (0.248) | (0.251) | (0.422) | (0.420) | (0.427) | (0.429) |
| Self-Evaluated Math Skills | -0.0750 | -0.0758 | -0.0767 | -0.0487 | -0.225 | -0.213 | -0.227 | -0.155 |
|  | (0.212) | (0.212) | (0.212) | (0.216) | (0.369) | (0.371) | (0.370) | (0.371) |
| Self-Evaluated Digital Skills | 0.296 | 0.302 | 0.309 | 0.306 | 0.703* | 0.612 | 0.715* | 0.729* |
|  | (0.196) | (0.201) | (0.201) | (0.198) | (0.399) | (0.417) | (0.399) | (0.414) |
| Constant | -1.231 | -1.224 | -1.266 | -1.338 | 5.229 | 5.118 | 5.194 | 4.940 |
|  | (1.712) | (1.708) | (1.730) | (1.741) | (3.245) | (3.257) | (3.262) | (3.100) |
|  |  |  |  |  |  |  |  |  |
| Model | OLS | OLS | OLS | OLS | OLS | OLS | OLS | OLS |
| Observations | 121 | 121 | 121 | 121 | 121 | 121 | 121 | 121 |
| R-squared | 0.271 | 0.271 | 0.273 | 0.276 | 0.308 | 0.312 | 0.309 | 0.320 |

Robust standard errors in parentheses
*** p<0.01, ** p<0.05, * p<0.1

Table 7: Supplemental Regression Results -- Linear Probability Model

| VARIABLES | (1) | (2) | (3) | (4) | (5) | (6) | (7) | (8) |
|---|---|---|---|---|---|---|---|---|
| | Writing Analysis | | | | Math & Data Analysis | | | |
| | Score Diff Dummy | Score Diff Dummy | Score Diff Dummy | Score Diff Dummy | Score Diff Dummy | Score Diff Dummy | Score Diff Dummy | Score Diff Dummy |
| ChatExpert30m | | -0.0294 | | | | 0.0531 | | |
| | | (0.0836) | | | | (0.0957) | | |
| ChatExpert1hr | | | -0.0384 | | | | -0.0282 | |
| | | | (0.0904) | | | | (0.103) | |
| ChatExpert2hr | | | | -0.137 | | | | -0.0217 |
| | | | | (0.110) | | | | (0.137) |
| Male | 0.146 | 0.144 | 0.144 | 0.150 | 0.0137 | 0.0172 | 0.0122 | 0.0143 |
| | (0.108) | (0.109) | (0.109) | (0.109) | (0.102) | (0.101) | (0.102) | (0.103) |
| Econometrics Grade | -0.167** | -0.165** | -0.168** | -0.177** | -0.0479 | -0.0510 | -0.0485 | -0.0496 |
| | (0.0657) | (0.0658) | (0.0667) | (0.0685) | (0.0800) | (0.0811) | (0.0799) | (0.0828) |
| GPAX | 0.170 | 0.163 | 0.173 | 0.180 | -0.0231 | -0.0106 | -0.0204 | -0.0215 |
| | (0.160) | (0.160) | (0.162) | (0.163) | (0.151) | (0.157) | (0.152) | (0.152) |
| Self-Evaluated Reading Skills | -0.0695 | -0.0714 | -0.0718 | -0.0725 | 0.00504 | 0.00848 | 0.00339 | 0.00456 |
| | (0.0595) | (0.0600) | (0.0597) | (0.0596) | (0.0611) | (0.0608) | (0.0623) | (0.0615) |
| Self-Evaluated Writing Skills | 0.00348 | 0.00246 | 0.00390 | 1.19e-05 | -0.0372 | -0.0354 | -0.0369 | -0.0378 |
| | (0.0569) | (0.0570) | (0.0571) | (0.0586) | (0.0620) | (0.0623) | (0.0624) | (0.0632) |
| Self-Evaluated Math Skills | -0.00444 | -0.00518 | -0.00475 | 0.00360 | -0.0325 | -0.0312 | -0.0328 | -0.0313 |
| | (0.0549) | (0.0551) | (0.0552) | (0.0560) | (0.0526) | (0.0533) | (0.0530) | (0.0530) |
| Self-Evaluated Digital Skills | 0.0749 | 0.0803* | 0.0772 | 0.0779 | 0.143** | 0.133** | 0.145** | 0.144** |
| | (0.0479) | (0.0475) | (0.0482) | (0.0480) | (0.0618) | (0.0667) | (0.0620) | (0.0629) |
| Constant | 0.583 | 0.589 | 0.576 | 0.550 | 0.568 | 0.556 | 0.563 | 0.562 |
| | (0.419) | (0.422) | (0.423) | (0.423) | (0.414) | (0.418) | (0.414) | (0.409) |
| | | | | | | | | |
| Model | LPM | LPM | LPM | LPM | LPM | LPM | LPM | LPM |
| Observations | 121 | 121 | 121 | 121 | 121 | 121 | 121 | 121 |
| R-squared | 0.211 | 0.212 | 0.212 | 0.220 | 0.244 | 0.246 | 0.244 | 0.244 |

Robust standard errors in parentheses
*** p<0.01, ** p<0.05, * p<0.1

Table 8: Supplemental Regression Results -- Logistic Model

|  | (1) | (2) | (3) | (4) | (5) | (6) | (7) | (8) |
|---|---|---|---|---|---|---|---|---|
|  | Writing Analysis | | | | Math & Data Analysis | | | |
| VARIABLES | Score Diff Dummy | Score Diff Dummy | Score Diff Dummy | Score Diff Dummy | Score Diff Dummy | Score Diff Dummy | Score Diff Dummy | Score Diff Dummy |
|  |  |  |  |  |  |  |  |  |
| ChatExpert30m |  | -0.0925 |  |  |  | 0.294 |  |  |
|  |  | (0.438) |  |  |  | (0.467) |  |  |
| ChatExpert1hr |  |  | -0.0714 |  |  |  | -0.109 |  |
|  |  |  | (0.520) |  |  |  | (0.534) |  |
| ChatExpert2hr |  |  |  | -0.810 |  |  |  | -0.129 |
|  |  |  |  | (0.682) |  |  |  | (0.701) |
| Male | 0.704 | 0.693 | 0.696 | 0.686 | 0.0550 | 0.0656 | 0.0504 | 0.0562 |
|  | (0.596) | (0.601) | (0.606) | (0.607) | (0.505) | (0.498) | (0.504) | (0.507) |
| Econometrics Grade | -1.051** | -1.047** | -1.048** | -1.098** | -0.264 | -0.279 | -0.265 | -0.272 |
|  | (0.429) | (0.430) | (0.428) | (0.454) | (0.420) | (0.420) | (0.420) | (0.427) |
| GPAX | 0.916 | 0.890 | 0.918 | 0.945 | -0.141 | -0.0662 | -0.132 | -0.135 |
|  | (0.876) | (0.880) | (0.881) | (0.904) | (0.752) | (0.789) | (0.751) | (0.749) |
| Self-Evaluated Reading Skills | -0.383 | -0.389 | -0.388 | -0.395 | 0.0654 | 0.0874 | 0.0587 | 0.0634 |
|  | (0.318) | (0.319) | (0.318) | (0.319) | (0.307) | (0.307) | (0.312) | (0.308) |
| Self-Evaluated Writing Skills | 0.0172 | 0.0133 | 0.0180 | -0.0108 | -0.244 | -0.230 | -0.245 | -0.250 |
|  | (0.313) | (0.312) | (0.312) | (0.321) | (0.319) | (0.321) | (0.321) | (0.328) |
| Self-Evaluated Math Skills | -0.0470 | -0.0498 | -0.0480 | -0.00198 | -0.196 | -0.196 | -0.195 | -0.187 |
|  | (0.289) | (0.289) | (0.290) | (0.295) | (0.273) | (0.275) | (0.272) | (0.273) |
| Self-Evaluated Digital Skills | 0.476* | 0.494* | 0.478* | 0.502* | 0.801** | 0.751* | 0.805** | 0.804** |
|  | (0.269) | (0.265) | (0.269) | (0.269) | (0.368) | (0.385) | (0.367) | (0.373) |
| Constant | 0.1000 | 0.205 | 0.119 | 0.252 | -1.023 | -1.333 | -1.005 | -1.007 |
|  | (2.512) | (2.585) | (2.506) | (2.588) | (2.270) | (2.359) | (2.269) | (2.280) |
|  |  |  |  |  |  |  |  |  |
| Model | Logistic | Logistic | Logistic | Logistic | Logistic | Logistic | Logistic | Logistic |
| Observations | 119 | 119 | 119 | 119 | 119 | 119 | 119 | 119 |
| R-squared |  |  |  |  |  |  |  |  |

Robust standard errors in parentheses
*** p<0.01, ** p<0.05, * p<0.1

Figure 1: Preliminary Results

Panel A: Writing Analysis Score
No ChatGPT vs. With ChatGPT

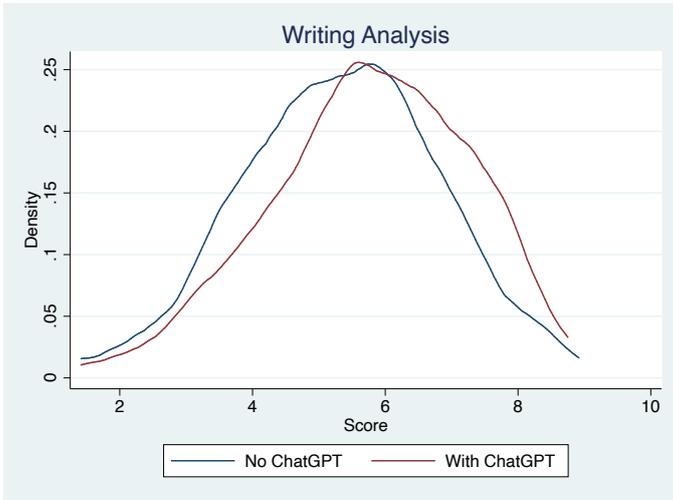

Panel B: Math & Data Analysis Score
No ChatGPT vs. With ChatGPT

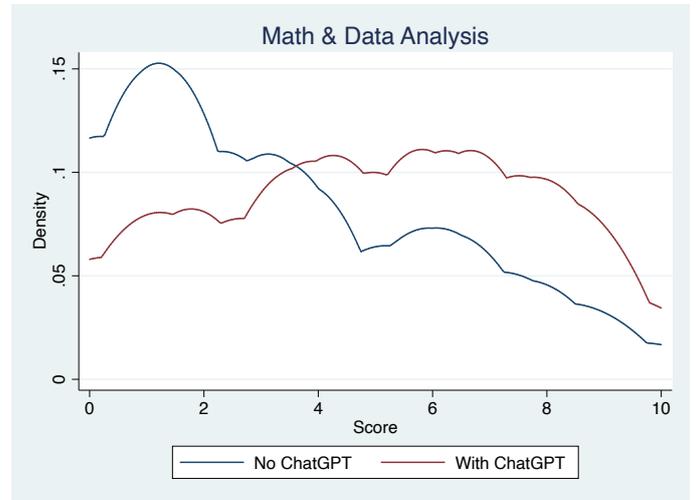

Panel C: Writing Analysis Time Spent
No ChatGPT vs. With ChatGPT

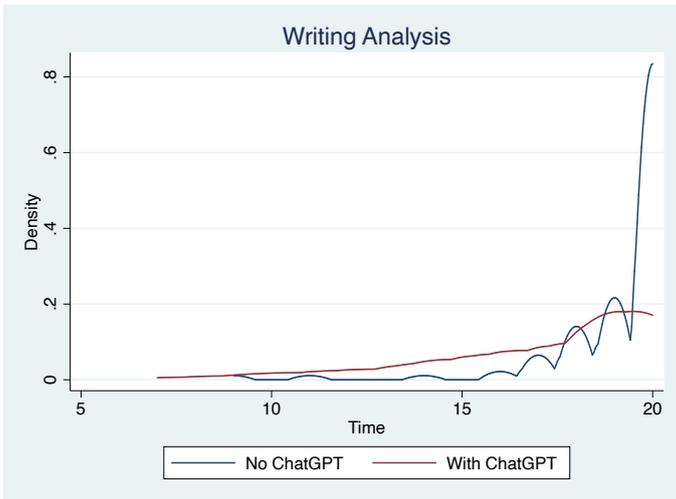

Panel D: Math & Data Analysis Time Spent
No ChatGPT vs. With ChatGPT

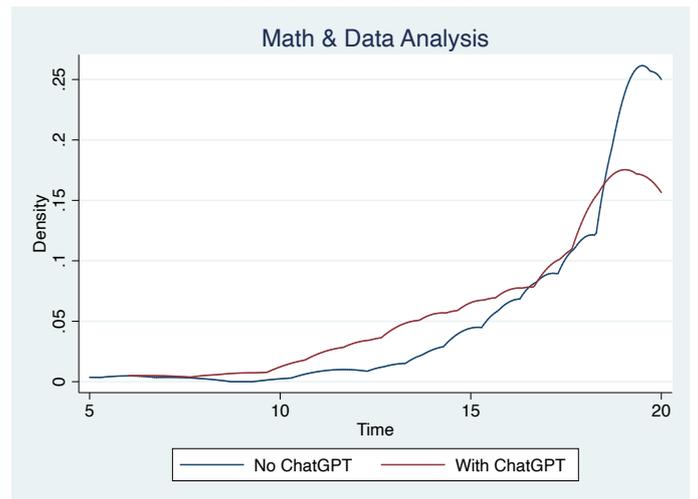

Figure 2: Detailed Results -- Writing Analysis

Panel A: Distribution of GPAX
Score Diff > 0 vs. Score Diff <= 0

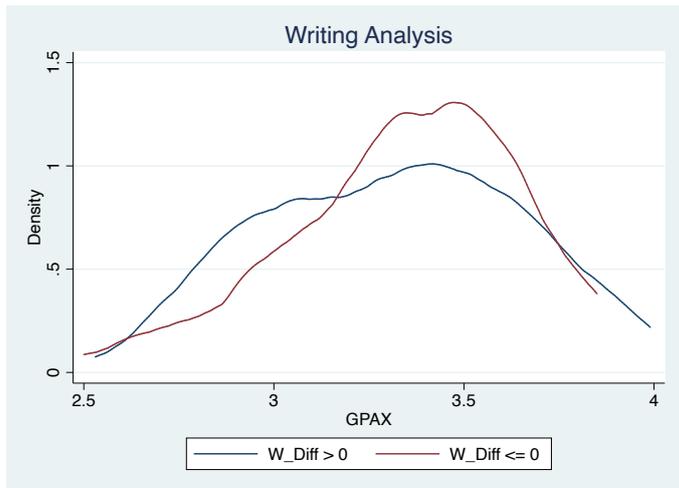

Panel B: Distribution of Econometrics Grade
Score Diff > 0 vs. Score Diff <= 0

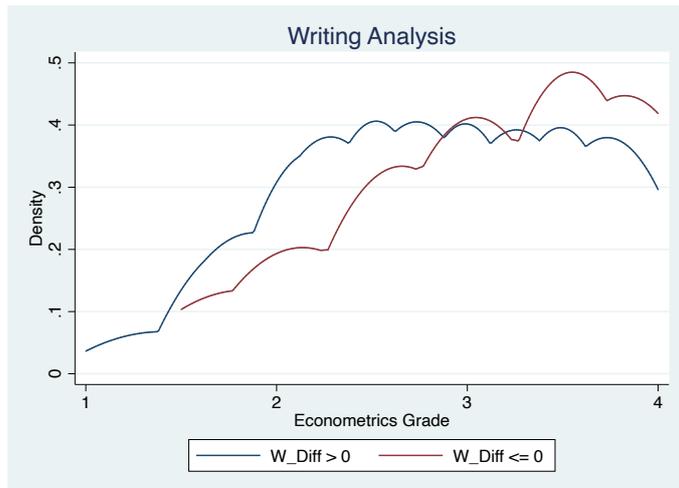

Panel C: Distribution of Self-Evaluated Reading Skills
Score Diff > 0 vs. Score Diff <= 0

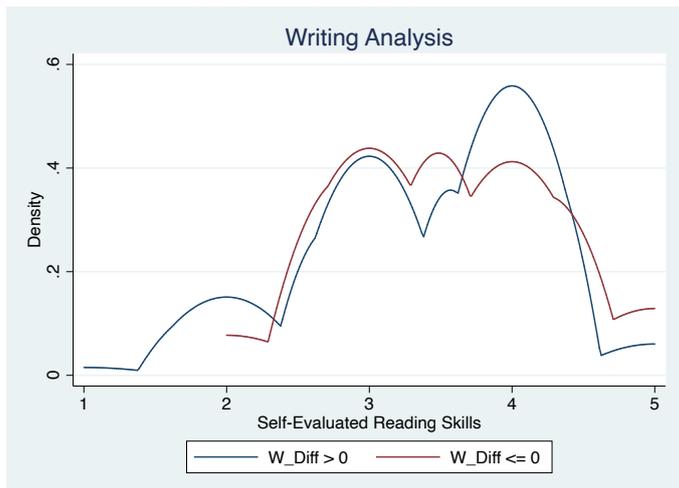

Panel D: Distribution of Self-Evaluated Writing Skills
Score Diff > 0 vs. Score Diff <= 0

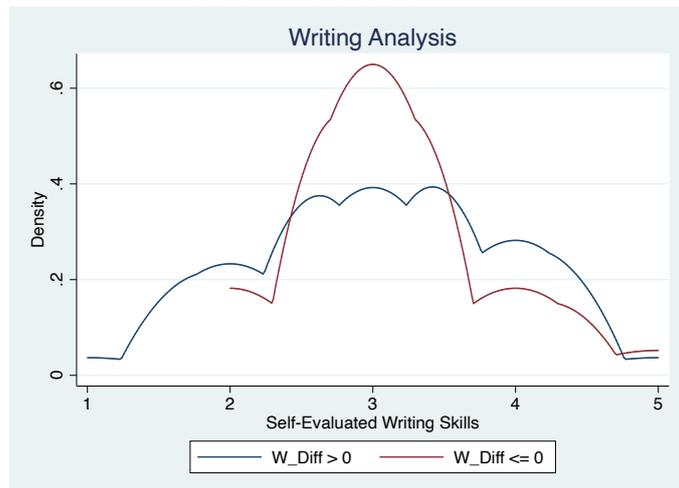

Panel E: Distribution of Self-Evaluated Math Skills
Score Diff > 0 vs. Score Diff <= 0

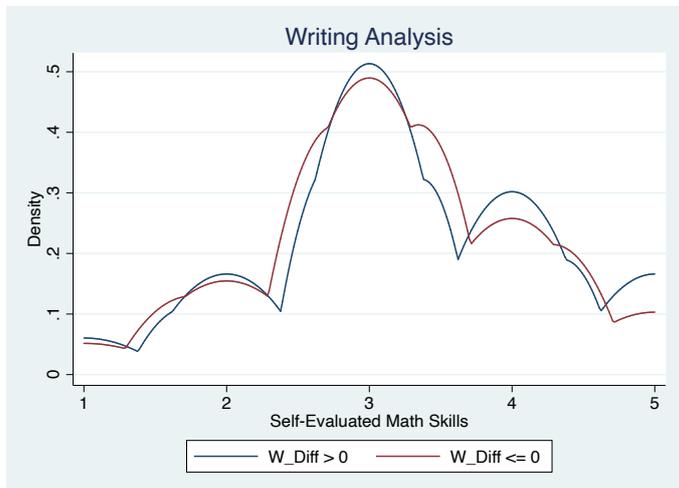

Panel F: Distribution of Self-Evaluated Digital Skills
Score Diff > 0 vs. Score Diff <= 0

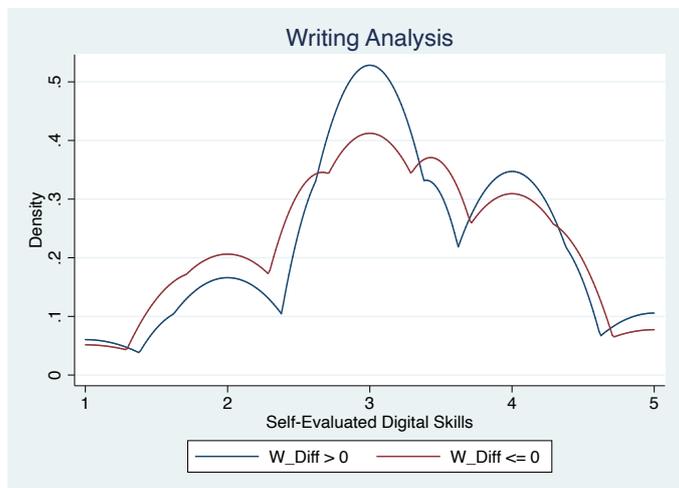

Figure 3: Detailed Results -- Math & Data Analysis

Panel A: Distribution of GPAX
Score Diff > 0 vs. Score Diff <= 0

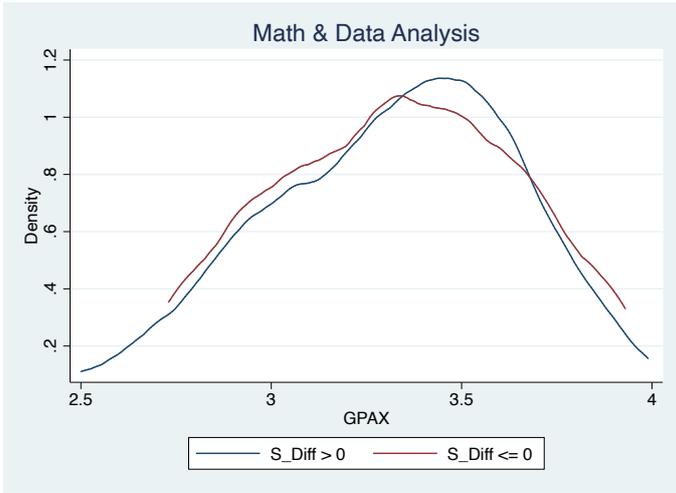

Panel B: Distribution of Econometrics Grade
Score Diff > 0 vs. Score Diff <= 0

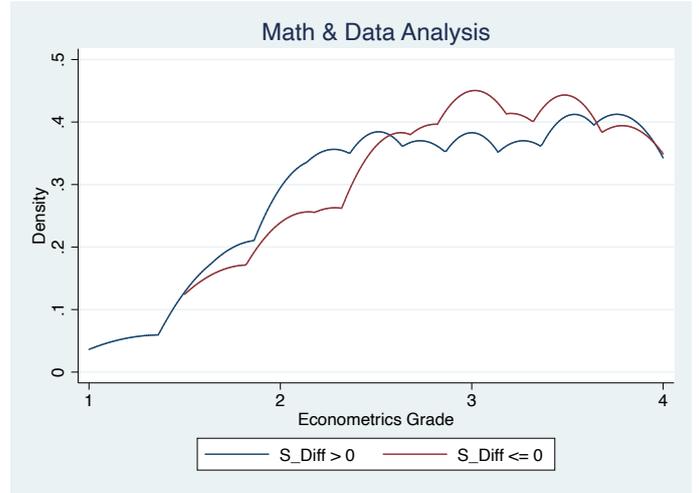

Panel C: Distribution of Self-Evaluated Reading Skills
Score Diff > 0 vs. Score Diff <= 0

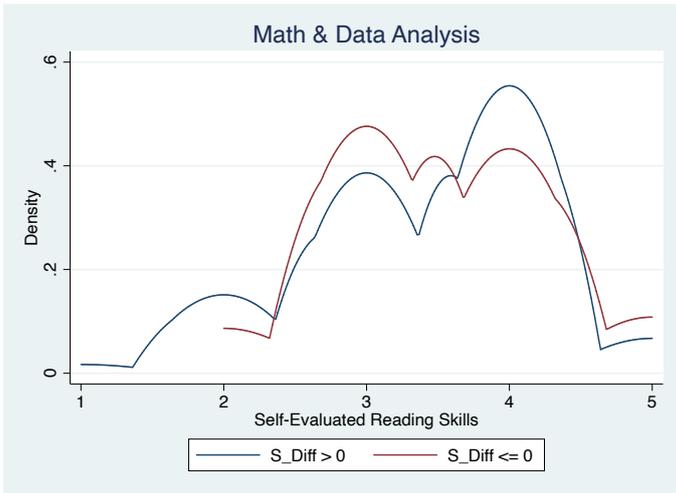

Panel D: Distribution of Self-Evaluated Writing Skills
Score Diff > 0 vs. Score Diff <= 0

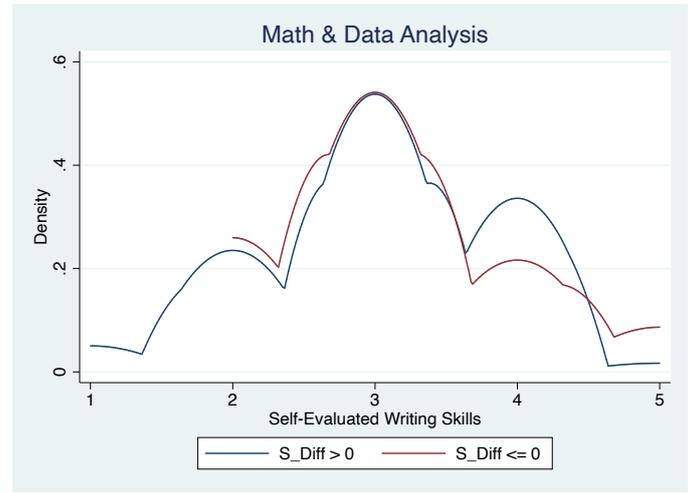

Panel E: Distribution of Self-Evaluated Math Skills
Score Diff > 0 vs. Score Diff <= 0

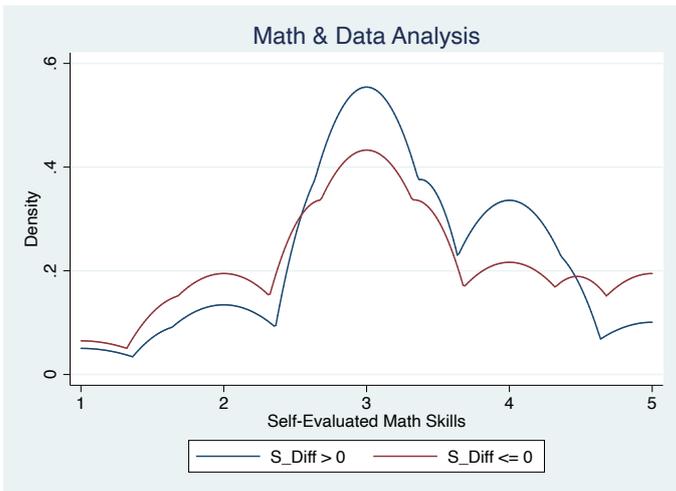

Panel F: Distribution of Self-Evaluated Digital Skills
Score Diff > 0 vs. Score Diff <= 0

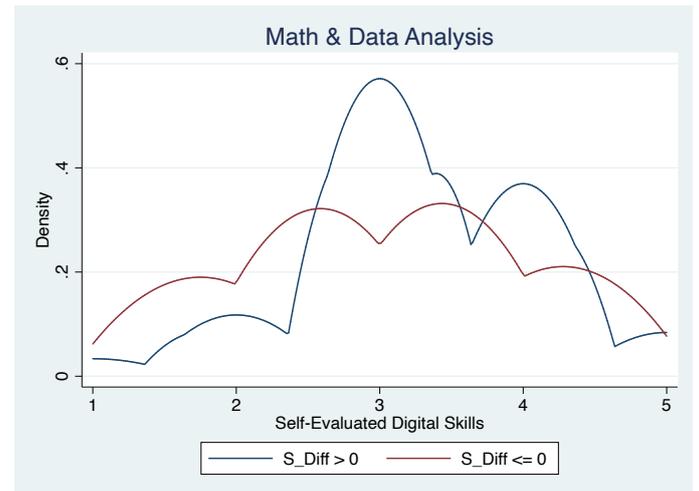